%% file: arxiv.tex
\documentclass{article}

\usepackage{publisher-inputs/arxiv/PRIMEarxiv}

\usepackage[utf8]{inputenc}
\usepackage[T1]{fontenc}
\usepackage{url}
\usepackage{booktabs}
\usepackage{amsfonts}
\usepackage{amsmath}
\usepackage{nicefrac}
\usepackage{microtype}
\usepackage{fancyhdr}
\usepackage{graphicx}
\usepackage{orcidlink}
\usepackage{amssymb}
\usepackage{algorithm}
\usepackage{algpseudocode}
\usepackage{tabularx}
\graphicspath{{media/}{figures/}}

\input{sections/title}

\author{
  Philip Huff\,\orcidlink{0000-0003-0869-2147} \\
  University of Arkansas at Little Rock \\
  \texttt{pdhuff@ualr.edu}
  \And
  Dakota Dale\,\orcidlink{0000-0002-4024-8030} \\
  Bastazo, Inc. \\
  \texttt{dakota@bastazo.com}
  \And
  Harshith Guduru\,\orcidlink{0009-0003-1302-9726} \\
  Bastazo, Inc. \\
  \texttt{harshith@bastazo.com}
  \And
  Rohan Singh\,\orcidlink{0000-0002-0309-1951} \\
  Bastazo, Inc. \\
  \texttt{rohan@bastazo.com}
  \And
  Qinghua Li\,\orcidlink{0000-0002-7734-6216} \\
  University of Arkansas \\
  \texttt{qinghual@uark.edu}
}
\date{v1: May 9th, 2026}

\begin{document}
\maketitle

\begin{abstract}
\input{sections/abstract}
\end{abstract}

\input{sections/keywords}

\input{sections/main}

\section*{Acknowledgments}
This material is based upon work supported by the U.S.\ Department of Energy under Award Number DE-CR0000031.

We thank the following research analysts for their substantial assistance with data collection and preliminary analysis for this study: Patrick Roberts, Nathan Thomason, Johnathan Reese,  and Carter Wallace.

\appendix

\section{Ethical Considerations}
This work focuses on defensive cybersecurity techniques, including mitigation planning and attack-path modeling. While the methodologies could be adapted for adversarial purposes, safeguards are taken to ensure responsible use. No human subjects or sensitive personal data were involved. The work aligns with responsible disclosure principles and aims to improve defensive capabilities for critical infrastructure.

\section{Open Science}
We provide an anonymous, self-contained repository containing all artifacts necessary to evaluate the core contributions of this work. The repository includes the full implementation of the mitigation planning pipeline, comprising the DRL environment and agent, adversary modeling via a VOMM, attack-path reconstruction using beam search, and budget-constrained mitigation optimization. It also includes pretrained models (DQN and VOMM), synthetic organizational datasets, configuration files, and scripts for executing each stage of the pipeline.

All artifacts are accessible through the repository at:
\begin{center}
\url{https://github.com/UALR-CORE-Center/strategic-mitigation-drl}
\end{center}
Reviewers can reproduce the full pipeline by installing dependencies and running a single entry-point script, which executes the end-to-end workflow and generates all outputs used in the paper, including reconstructed attack paths, mitigation plans, and figures. The demo executes in seconds on a standard laptop and does not require external services, credentials, or network access.

To support reproducibility, all randomness is seeded, required data is included locally, and pretrained models are provided to avoid the need for expensive retraining. The environment also supports deterministic simulation through explicit state capture and restoration to enable consistent evaluation of attack-path reconstruction and mitigation decisions.

Certain upstream artifacts used in constructing the adversary models, specifically the full corpus of threat intelligence reports, cannot be redistributed due to licensing and aggregation constraints. To address this, we provide derived adversary profiles, representative ATT\&CK datasets, and pretrained VOMM models that preserve the statistical properties required to evaluate the methodology.

These artifacts are sufficient to reproduce and assess all primary contributions of the paper, including the constrained MDP formulation, integration of adversary modeling with reinforcement learning, attack-path reconstruction, and budget-aware mitigation planning.

\input{sections/appendices}

\bibliographystyle{unsrt}
\bibliography{sections/references}

\end{document}

%% file: sections/title.tex
\title{Operationalizing Cybersecurity Governance for Mitigation Planning with Attack-Path Modeling and Reinforcement Learning}

%% file: sections/abstract.tex
We address a fundamental challenge in cybersecurity operations of translating governance frameworks into actionable mitigation decisions under realistic resource constraints. Frameworks such as the NIST Cybersecurity Framework (CSF) provide widely adopted measures of organizational maturity, but do not directly support the selection and prioritization of defensive strategies against adversarial behavior. We present a system that operationalizes governance frameworks by mapping CSF maturity assessments into MITRE ATT\&CK mitigation capabilities, which enables direct integration of organizational security posture with adversary-informed defensive planning.

To manage adversary complexity, we employ a Variable-Order Markov Model (VOMM) trained on observed ATT\&CK technique sequences to enable scalable adversary simulation within a Deep Reinforcement Learning (DRL) environment. We reconstruct likely attack paths and defensive responses using beam search, and then jointly optimize mitigation selection under explicit budget constraints. 

Our environment supports concurrent adversaries and realistic mitigation costs. Across multiple reward formulations and configurations, we show that the approach produces stable policies, meaningful cost–risk trade-offs, and interpretable mitigation plans aligned with organizational maturity. These results demonstrate that adversary-aware DRL can generate practical, resource-constrained defense strategies grounded in real-world frameworks and threat behavior.

%% file: sections/keywords.tex
\keywords{Cyber Defense \and Deep Reinforcement Learning \and Attack-Path Modeling \and Automated Mitigation Planning \and Markov Decision Processes}

%% file: sections/main.tex
\section{Introduction}
Cyber defenders must continuously allocate scarce people, time, and budget to withstand a changing landscape of adversarial tactics, techniques, and procedures (TTPs). Control catalogs such as the NIST Cybersecurity Framework (CSF) and NIST SP 800-53 offer comprehensive guidance, yet organizations routinely struggle to translate these abstractions into actionable mitigation work on specific systems \cite{jiang2025mitre}. In practice, the obstacles are threefold: (i) bridging the gap between high-level controls and system-level actions, (ii) planning defensive countermeasures at scale under real operational constraints, and (iii) keeping pace with adversaries who adapt their TTPs faster than defenses can be reconfigured. Because defenders cannot protect everything at all times, the problem is one of prioritization and optimization. Similarly, as observed in recent industry analysis, adversaries tend to reuse a relatively small set of techniques rather than introducing fundamentally novel tradecraft \cite{donohue2025forevertechniques}, suggesting that decision support systems tuned to likely adversarial behavior can yield materially better defensive outcomes.

Recent research has demonstrated the value of reinforcement learning (RL) for cyber defense through interactive game environments such as CybORG and its CAGE variants, which emphasize autonomous real-time defensive actions against adaptive adversaries \cite{standen2021cyborg, kiely2025cage}. These benchmarks have been critical in advancing cyber defense agents and standardizing evaluation under adversarial uncertainty. However, this line of work is optimized for reactive, short-horizon operational actions (e.g., isolating a host, restoring a service) and does not explicitly model portfolio-style mitigation planning under organizational maturity and budget constraints. In practice, security teams must translate threat intelligence and control frameworks into feasible mitigation plans that align with available resources and that can be executed over normal planning cadences such as over a monthly sprint. This gap motivates a complementary RL formulation centered on strategic mitigation planning, where defender actions correspond to realistic mitigation decisions grounded in an organization's capability and threat profile.

We present a learning environment that couples observed adversary behavior with organizational mitigation planning. Our objective is not autonomous deployment of countermeasures on production systems, but rather the generation of cost-effective mitigation portfolios that human defenders can validate and implement. To bridge organizational security practices with operational mitigations, we use LLMs to infer supportive relationships between NIST CSF practices and MITRE ATT\&CK mitigations to enable commonly collected CSF practice assessments to be translated into mitigation maturity levels. Then, to preserve realism while avoiding the collection of private organizational data, we synthesize heterogeneous organizational maturity profiles into plausible distributions of defender capability. Finally, to ground adversary behavior in observed tradecraft without enumerating the full ATT\&CK technique space, we use an empirical prior for adversary technique sequencing during simulation.

\begin{figure}[t]
  \centering
  \includegraphics[width=\linewidth]{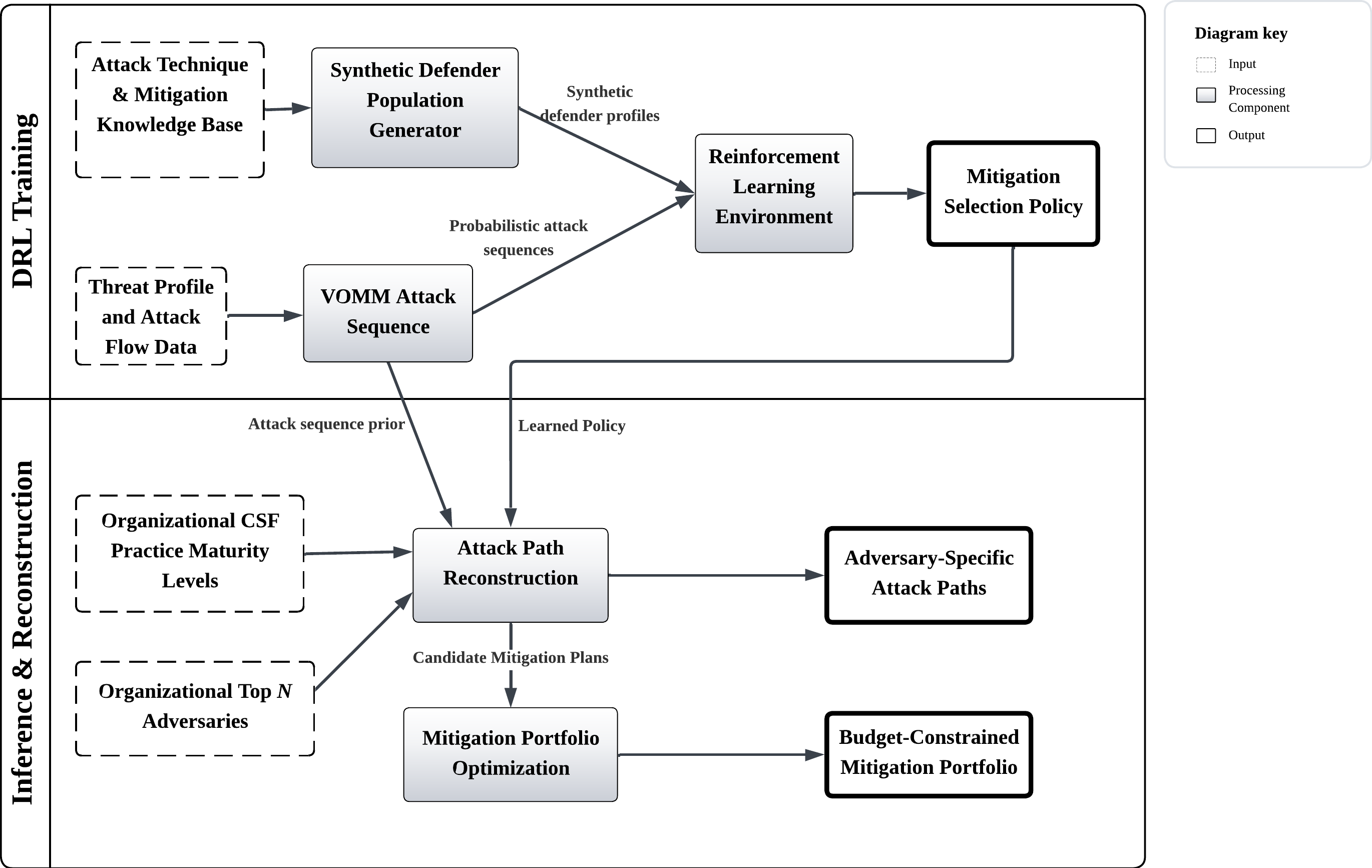}
  \caption{%
  System overview of our strategic mitigation planning framework.
  During training (top), organizational knowledge and threat intelligence are used to generate synthetic defender populations and probabilistic adversary behavior models, which are integrated into a reinforcement learning environment to learn mitigation selection policies under explicit cost constraints.
  During inference (bottom), organizational maturity assessments and a selected set of relevant adversaries are combined with learned policies to reconstruct likely attack paths and identify budget-constrained mitigation portfolios.
  The architecture separates offline learning from online planning and produces interpretable mitigation recommendations used for operational decision support.
  }
  \label{fig:system-overview}
\end{figure}

Our environment models cyber defense on specific systems within an organization against its most relevant adversaries across discrete planning periods aligned with normal operational cadence. At the start of each episode, the defender observes its mitigation maturity profile and a set of likely adversaries, then selects a budget-feasible portfolio of mitigations. The portfolio remains fixed while one or more adversaries attempt to progress along sampled ATT\&CK technique sequences, with success probabilities shaped by adversary capability and reduced by the selected mitigations as a function of organizational mitigation maturity. The reward structure captures mission-centric outcomes (preventing adversary impact) and operational efficiency (risk reduction per unit cost). After training, we reconstruct high-likelihood attack and defense paths to provide traceability linking recommended mitigations to the specific adversary techniques they are intended to disrupt. Finally, we select a budget-feasible subset of these candidate work items for a given planning period using explicit organizational budget constraints. Figure~\ref{fig:system-overview} illustrates the overall architecture of our approach and the separation between offline learning and online mitigation planning.

We make four contributions. (1) We formulate strategic mitigation planning as a constrained Markov decision process (MDP) in which actions correspond to portfolios of mitigations that must satisfy explicit cost constraints, rather than to atomic defensive responses. Unlike prior cybersecurity MDP and reinforcement learning formulations that emphasize reactive containment or single adversaries, our environment supports concurrent adversary campaigns and long-term planning over organizational mitigation investments. (2) We introduce an LLM-assisted semantic translation layer that converts NIST CSF practice assessments into mitigation maturity levels, bridging a long-standing abstraction gap between governance-oriented frameworks and operational defense actions. We pair this translation with an ordered logit population generator that produces statistically diverse and realistic organizational maturity profiles, enabling scalable training and controlled evaluation across heterogeneous defenders. (3) We incorporate an empirical adversary behavior prior learned from ATT\&CK technique sequences to constrain adversary progression, improving both simulation realism and sample efficiency during training. Building on this structure, we reconstruct likely attack and defense paths to produce interpretable mitigation recommendations that explicitly account for budget constraints and are suitable for operational planning rather than generic policy outputs. (4) We identify a budget-constrained 0-1 knapsack optimization that converts reconstructed attack-path insights into an actionable mitigation plan.

The remainder of the paper proceeds as follows. Section~\ref{sec:env-setup} defines our environment, including organizational maturity and mitigation feasibility, synthetic population generation, attack dynamics and sequencing, and adversary modeling from threat intelligence. Section~\ref{sec:decision-formulation} formalizes the induced MDP, including observation and action spaces, transition dynamics, and reward and cost modeling. We then describe attack-path reconstruction and analysis for interpretability in Section~\ref{sec:attack_path_reconstruction}, followed by Section~\ref{sec:experiment} with our experimental evaluation and results, real-world validation, and a discussion of limitations and future work.

\section{Background}

\subsection{MITRE ATT\&CK Concepts}
The MITRE ATT\&CK framework was introduced to systematically enumerate the tactics and techniques observed in real-world cyber adversary behavior and to support the operationalization of threat intelligence for defenders \cite{strom2018attack}.

The current ATT\&CK Enterprise matrix organizes adversary behavior into thirteen high-level tactics representing a typical progression of an attack, from reconnaissance through impact. Each tactic comprises a set of techniques that describe specific adversarial actions observed in the wild or anticipated based on known capabilities \cite{mitreICS}.

ATT\&CK techniques are further mapped to mitigations that represent defensive actions organizations can implement to reduce the likelihood or impact of adversary activity. While these mappings provide valuable conceptual guidance, the overall ATT\&CK model space remains large with approximately 900 adversarial techniques and over one hundred mitigation categories. This scale makes direct incorporation of the full ATT\&CK action space in RL environments computationally impractical and misaligned with how defenders plan and execute mitigation work.

To address this challenge, we leverage ATT\&CK attack flows as a means of constraining adversary behavior to realistic paths \cite{mitreAttackFlow}. Attack flows are structured representations of empirically observed sequences of adversary techniques. Prior analyses have shown that real-world intrusion sets exhibit substantial overlap in the tactics and techniques they employ \cite{donohue2025forevertechniques}. By using existing attack flows, we retain empirical fidelity while substantially reducing the effective action space of the RL environment.

These attack flows also highlight an important modeling challenge that adversary behavior in the ATT\&CK framework is inherently sequential and path-dependent. However, empirical data on complete attack sequences remains sparse and unevenly distributed across techniques. Learning a full adversary policy directly within a reinforcement learning environment would therefore require extensive exploration and large training corpora. Instead, we employ a Variable-Order Markov Model (VOMM) as a lightweight, non-parametric probabilistic reasoning model that captures recurring mid-range dependencies in adversary technique sequences. The VOMM provides a probabilistic bias toward empirically observed transitions that shapes exploration while allowing the learning agent to deviate when rewards favor novel paths.

\subsection{NIST CSF 2.0 Practices and Maturity Tiers}
The NIST Cybersecurity Framework (CSF) was developed in response to Executive Order 13636, Improving Critical Infrastructure Cybersecurity, to provide a flexible, risk-based approach for managing cybersecurity risk across heterogeneous organizations and sectors \cite{eo13636,nistcsf2}. Rather than prescribing specific security controls, the CSF defines a set of outcome-oriented cybersecurity practices that better aid organizations in adapting implementation strategies based on mission, risk tolerance, and available resources. This design distinguishes the CSF from control frameworks such as NIST SP 800-53. In this work, we use CSF practice maturity as a proxy for the feasibility and cost of applying mitigation recommendations.

Cybersecurity mitigations are executed within organizational and system constraints that are largely determined by the availability of tools, automation, and skilled personnel. Prior work has shown that insufficient capital investment in foundational cybersecurity assets forces organizations to rely disproportionately on recurring operational effort to compensate for missing preventative controls \cite{brho2025finance}. Within the context of the CSF, limited capital investment constrains achievable practice maturity.

By grounding our cost model in CSF practice maturity, we estimate the operational costs for organizations to apply mitigation. This abstraction allows cost–benefit optimization that is tailored to an organization’s existing capabilities.

\section{Related Work}
Early work on cyber defense optimization has framed mitigation as a sequential decision-making problem under uncertainty using a Markov Decision Processes (MDPs) and Partially Observable Markov Decision Processes (POMDPs). In these formulations, the defender selects countermeasures in response to partially observed attacker behavior, explicitly modeling noisy alerts and hidden attacker state \cite{durkota2015optimal, applebaum2016intelligent, horak2017heuristic, panfili.2018}. Defender actions in these models typically include a small set of operational responses such as isolating hosts, deploying honeypots, or restricting access. Purves et al.\ integrate Structural Causal Models into PPO by explicitly modeling reward dynamics to improve both interpretability and performance in cyber defense environments \cite{purves2024causally}.

Recent survey work highlights that, while these approaches provide strong theoretical foundations, they face scalability challenges in realistic environments, where multi-stage attacks, heterogeneous assets, and expanding mitigation options cause the state and action spaces to grow rapidly \cite{ozkan2024comprehensive}. This leaves open challenges in scaling to realistic environments and supporting strategic, resource-constrained planning.

\subsection{RL Cyber Defense Environments and Benchmarks}
A growing body of work focuses on building RL testbeds and examining how to make DRL-based cyber defense agents more interpretable and operationally grounded. Castro et al.\ use an LLM agent to interpret signals from the CybORG CAGE environment and issue responses, which demonstrates that LLMs can help explain and contextualize agent behavior, in contrast to DRL agents that act quickly but are difficult to interpret \cite{Castro2025-ga}. Loevenich et al.\ similarly address interpretability limitations, but instead augment an autonomous DRL agent with a hierarchy of cybersecurity knowledge graphs and an LLM analytic layer that synthesizes logs, threat intelligence, and system context to explain alerts and defensive decisions \cite{Loevenich2025-st}.

Mukherjee et al.\ introduce an LLM-assisted reward design framework in which a commercial LLM generates reward structures for both attacker and defender personas in the Cyberwheel simulation environment \cite{Mukherjee2025-ob}. These LLM-derived rewards are then used to train DRL defenders, showing that LLM-based reward engineering can shape defender behavior and yield more effective strategies across varied adversarial profiles. Oesch et al.\ take a broader systems-level perspective, arguing that progress toward autonomous cyber defense requires specialized multi-agent architectures aligned with the NIST Cybersecurity Framework (CSF) functions, and emphasizing two central challenges for DRL in this setting, which includes defining the right game (observation spaces, action spaces, detectors, and rewards that match operational reality) and enabling agents to adapt to changing networks, missions, and attacker strategies \cite{Oesch2025-ci}.

Complementary work demonstrates the potential of DRL itself in more realistic environments. Thompson et al.\ show that entity-based reinforcement learning with transformer policies substantially improves the generalization of autonomous cyber defense agents across variable network topologies \cite{symes2023entity}. Yu et al.\ demonstrate that DRL can learn robust and adaptive defense policies under partial observability and noisy telemetry, which points to establishing the feasibility of autonomous defense agents in realistic cyber environments \cite{yu2023airs}.

Recent work has also begun to systematically evaluate multiple DRL algorithms within unified cyber defense environments. Hammad and Jasim present a comparative study of five DRL approaches: DQN, PPO, TD3, A3C, and SAC, and evaluate in a realistic network simulation with live-streaming traffic \cite{hammad2025adaptive}. Their results show that entropy-regularized actor-critic methods, particularly SAC, consistently outperform value-based approaches in detection accuracy, adaptability, and operational efficiency. Wilson et al.\ extend this line of work to operational technology (OT) environments, introducing a multi-agent RL testbed (IPMSRL) for industrial control systems and demonstrating that coordinated MARL policies (e.g., MAPPO) outperform independent learners while remaining robust under partial observability and imperfect alerting \cite{wilson2024multi}.

Despite these advances, most existing environments remain focused on real-time operational response, where agents execute defensive actions during ongoing attacks. This includes CybORG, CyberBattleSim, and related frameworks, which typically model simplified state representations and constrained action spaces. In contrast, our work targets a different phase of the cybersecurity lifecycle with strategic mitigation planning. Rather than assuming autonomous deployment of countermeasures in live systems, we use DRL to derive optimized mitigation strategies under budget constraints, with human defenders remaining in the loop for execution.

Our work differs from prior approaches in several important respects. First, whereas benchmark environments often provide abstracted representations with limited action spaces, we construct an environment grounded in operationally realistic data and expose a richer mitigation decision space that captures cost, maturity, and effectiveness trade-offs. Second, our objective is fundamentally distinct. Existing DRL cyber defense systems primarily optimize short-horizon, reactive decisions, while we focus on long-horizon planning under resource constraints.

Finally, in alignment with the research gaps identified by \cite{Oesch2025-ci}, our work contributes to defining a more generalized and operationally grounded decision framework for defender agents. Our environment, observation space, and attack-path reconstruction jointly anchor the DRL agent’s learning in realistic defensive contexts. Although we do not claim to solve the broader challenge of adaptability, we incorporate dynamic features such as variable adversarial attributes and mitigation maturity levels, enabling the learned policy to generalize more effectively across protection scenarios.

\subsection{Modeling Adversary Technique Sequences}
A small but growing literature explicitly models sequences of adversary techniques using MITRE ATT\&CK. Choi et al.\ propose a probabilistic framework based on a hidden Markov model to infer attack sequences from observed telemetry \cite{choi2021probabilistic}. Ahmed et al.\ use the MITRE ATT\&CK database to construct probabilistic attack graphs that feed into quantitative risk assessments \cite{ahmed2022mitre}. Kuwano et al.\ propose a recommendation-based approach for forecasting adversary behavior over the MITRE ATT\&CK framework by treating attacker groups as users and techniques as items, applying collaborative filtering to predict likely next techniques from partially observed attack activity \cite{kuwano2022att}. Inspired by this line of work, we construct a Variable Order Markov Model (VOMM) over observed cyber attacks derived from ATT\&CK techniques, which allows us to reduce the effective training space for RL while preserving realistic multi-step adversary behavior.

\subsection{LLMs for Cyber Defense}
Recent advances show that LLMs can reliably extract cyber threat intelligence into structured formats, including representations aligned with MITRE ATT\&CK and STIX-like schemas \cite{massengale2024linking, kravsovec2025large, microsoftopenai2024threatai}. We build on this capability to automatically map and score large portions of the ATT\&CK framework and to profile adversaries in terms of capability, cost, and impact. These structured representations serve as inputs to our environment design and risk modeling, enabling our DRL agents to plan mitigations against empirically grounded adversary behaviors.

\section{Environment Setup}
\label{sec:env-setup}
We now introduce our environment for strategic mitigation planning, which formalizes organizational defensive state, adversary behavior, and their interaction.

\subsection{Defender Organizational State and Mitigation Maturity}
\label{subsec-csf-mapping}
For the purpose of preparing our defender environment, we focus only on NIST CSF practices that serve as direct countermeasures to an adversary’s attack techniques. We select CSF practices because they are already associated with well-defined maturity levels and established assessment methods, which organizations often maintain as part of routine cybersecurity evaluations or can reasonably assess with limited additional effort. We exclude practices primarily reflecting governance and incident response processes, and focus instead on practices that translate into direct, actionable mitigations. This filtering results in a set of 42 practices that we use to characterize an organization’s mitigation maturity.

To model organizational maturity in the environment, we represent each organization by a maturity level that determines the resources available to implement mitigations. However, the MITRE ATT\&CK framework, which defines 95 enterprise and ICS mitigation techniques, is not designed to assess maturity directly. To bridge this gap, we construct a set-wise mapping between NIST CSF practices and ATT\&CK mitigations, where each CSF practice may support multiple mitigations and each mitigation may be supported by multiple practices. For example, the CSF practice PR.AA-01 (“identities and credentials are managed”) maps to ATT\&CK mitigations such as M1035 (Limit Access to Resource Over Network) and M1026 (Privileged Account Management), since the practice supports implementation of constraining remote reachability and privileged access paths.

Each practice–mitigation pair is assigned an ordinal strength score from 1 (very low) to 5 (very high) based on a fixed rubric that considers the directness with which the practice supports the mitigation’s mechanism. A score of 5 indicates that the mitigation cannot realistically be implemented without the practice, whereas a score of 3 indicates the practice enhances the effectiveness, consistency, or scalability of the mitigation, and a score of 1 means no meaningful relationship exists between the practice and the mitigation. We use a large language model (OpenAI o4-mini-2025-04-16) to apply this rubric at scale, as strength assignment is primarily a semantic alignment problem between two frameworks with differing abstractions and terminology. The model is provided with the official textual definitions of both CSF practices and ATT\&CK mitigations and instructed to produce rubric-consistent scores with brief rationales.

Rather than using a simple weighted average, we compute mitigation maturity using a nonlinear weighted power mean to attenuate weak relationships and emphasize strong ones. Relation strengths are first normalized to $[0,1]$. We then apply the weighted power mean aggregation:

\begin{equation}
\label{eq:mitigation_maturity_score}
\text{Mitigation\_Maturity}
=
\left(
\frac{
\sum_i w_i \cdot (\text{practice}_i \cdot w_i)^q
}{
\sum_i w_i
}
\right)^{\frac{1}{q}}
\end{equation}

where $w_i$ is the normalized relation strength and $\text{practice}_i$ is the practice maturity level. The constant $q > 1$ emphasizes stronger contributors over weaker ones. By incorporating $w_i$ both as a weight and within the nonlinear term, weakly related practices are sharply down-weighted while strongly supported relationships dominate the resulting score. This formulation increases dispersion in mitigation maturity scores and prevents convergence toward uniform averages across mitigations. The resulting mitigation maturity scores are then normalized to the interval $[0,1]$ to provide a realistic input signal for the DRL environment.

\subsection{Synthetic Data Generation Framework}
\label{subsec:synthetic-generation}
Training a DLR agent to reason over organizational cybersecurity requires a large collection of maturity profiles that are both realistic and representative of the lower maturity levels where most organizations reside. Because real-world datasets are constrained by privacy, legal, and regulatory considerations, we generate synthetic NIST CSF maturity profiles that preserve plausible cross-practice dependencies while biasing prevalence toward lower maturity. Each organization is assigned a latent maturity level drawn from a categorical prior that favors lower maturity, with a small stochastic perturbation to induce variability. Practice-level maturity tiers are then sampled using an ordered logit model with practice-specific difficulty adjustments derived from cost and complexity, ensuring that higher latent maturity increases the likelihood of higher tiers while preserving variation across practices.

The resulting CSF practice maturity tiers are mapped to ATT\&CK mitigation maturity signals using the weighted scheme described in Section~\ref{subsec-csf-mapping}. The full sampling procedure and parameterization are provided in Appendix~\ref{app:synthetic_data} and reproducible in our open science repository.

\subsection{Attack dynamics and sequencing}
\label{subsec-vomm}
With hundreds of ATT\&CK techniques and highly variable implementations, mapping techniques directly into a reinforcement learning action space leads to an exploration problem that is sparse and inefficient, often resulting in excessive training time. Rather than forcing the agent to learn attack-sequence structure from scratch or limiting the action space, we provide a behavioral prior learned from documented adversary operations. Specifically, we train a Variable-Order Markov Model (VOMM) on ATT\&CK Flow data so the agent begins with a distribution over what attackers are likely to do next, which guides exploration toward sequences that align with real-world adversary behaviors.

The VOMM conditions on the most recent attacker techniques and predicts a distribution over the next technique token. For example, given the partial path
\texttt{TA0001:T1190} (exploit public-facing application) $\rightarrow$ \texttt{TA0006:T1136.001} (create local account),
the model should assign higher probability to \texttt{TA0003:T1059} (command and scripting interpreter) than to an unrelated step such as \texttt{TA0040:T1490} (inhibit system recovery). The model is \emph{variable-order}: it uses the \emph{longest} available context up to length $K$ that is sufficiently supported by data; otherwise it backs off to shorter contexts.

We estimate transition probabilities using weighted counts with add-$\alpha$ smoothing and back-off:
\begin{equation}
\label{eq:vomm_smoothing_short}
p(a \mid \mathbf{c})
=
\frac{
C(\mathbf{c}, a) + \alpha
}{
C(\mathbf{c}) + \alpha |V|
}
\end{equation}
Full corpus construction, weighting, and back-off procedures are provided in Appendix~\ref{app:synthetic_data}.

\subsection{Adversary modeling from threat intelligence}
\label{subsec:adversary-modeling}
To construct realistic adversary models, we derive adversary-specific behaviors and cost characteristics from open-source threat intelligence reports using a multi-stage LLM extraction pipeline with human analyst validation. This process converts heterogeneous, unstructured reports into structured adversary profiles suitable for reinforcement learning. This is critical since effective defense requires not only identifying relevant mitigations, but instantiating them in a manner that directly corresponds to how adversaries operate in practice.

The corpus consists of approximately 1,900 publicly available threat intelligence reports published by government agencies and commercial security vendors, including CISA, Microsoft Security, Google Threat Analysis Group, Palo Alto Networks Unit 42, Dark Reading, and similar sources. Reports are ingested through a structured review workflow in a university cybersecurity clinic, where each report is independently reviewed by two trained analysts prior to final feature publication.

Across approximately 1,900 open source reports, we identify and model roughly 180 distinct adversaries. Each adversary is represented as a set of profiling characeristics and observed ATT\&CK techniques. Rather than treating techniques as abstract identifiers, we model them as concrete actions grounded in reported attacker behavior, which enable the derivation of mitigation strategies directly tied to how adversaries execute attacks in practice.

In the first stage, the LLM extracts all explicitly observed ATT\&CK techniques from each report along with textual descriptions of how the adversary executes them. For each technique instance, we assign a five-level Likert score capturing the cost and complexity incurred by the adversary. All extracted techniques and effort estimates are reviewed and validated by human analysts before inclusion.

In the second stage, each validated technique instance is mapped to candidate ATT\&CK mitigations using the official MITRE ATT\&CK technique–mitigation mappings as a baseline. The LLM is not used to invent new technique–mitigation relationships. Instead, it refines mitigation instances by conditioning on the adversary’s observed execution details and provides contextual instantiations of existing mitigations (e.g., tailoring network controls to protocol, tooling, or infrastructure described in the report). For each mitigation instance, the LLM assigns five-level Likert scores for defender implementation difficulty, expected effectiveness, and cost. These are also subject to human analyst validation. This produces a direct linkage between concrete adversary actions and actionable defensive responses, and ensures that mitigation recommendations are not generic controls but are instead grounded in the specific techniques and execution patterns observed in real attacks.

To align adversary techniques with a given defender organization, we identify each organization’s most likely adversaries using profile similarity methods from \cite{massengale2024linking,massengale2024assessing}. This allows attack paths and mitigation decisions to be evaluated against adversaries that are relevant to the organization, rather than against a generic global threat model.

\paragraph{Example.} The ATT\&CK technique 
\texttt{T1041} (Exfiltration Over C2 Channel) 
and corresponding 
\texttt{M1031} (Network Intrusion Prevention) 
are highly abstract and unusable in isolation for defender modeling. However, in a representative report, the pipeline extracts a concrete behavior of custom PowerShell scripts  \texttt{lootsubmit.ps1} and \texttt{trackerjacker.ps1} exfiltrating system and location data via HTTPS POST requests to a \texttt{netlify.app} endpoint. 

From this description, the system derives a contextualized mitigation instance that specifies the class of defensive actions required (e.g., outbound HTTPS inspection, signature-based detection of script behavior, and filtering of known destination infrastructure). The pipeline does not generate specific organizational rules, such as IP addresses or firewall policies. Instead, it produces structured mitigation templates and candidate detection artifacts (e.g., YARA, Sigma, or Elastic rule patterns) that parameterize defender effort and expected effectiveness without assuming deployment details.

The final output of this pipeline is a set of adversary profiles comprising (i) technique sequences used during attack simulation, (ii) adversary effort parameters influencing attack dynamics, and (iii) mitigation instances with defender-specific cost and effectiveness attributes. These profiles parameterize both attack-path reconstruction and reward–cost trade-offs in the DRL environment to ensure that learned mitigation strategies are optimized against realistic threats and grounded in the specific adversary behaviors they are intended to disrupt.

\section{Decision Formulation}
\label{sec:decision-formulation}
We cast mitigation planning as a sequential decision problem in which a defender allocates limited effort to reduce the feasibility of likely adversarial TTPs. This section specifies the observation space, action space, transition dynamics, and reward function used to learn mitigation policies.

The defender observes (i) an estimated distribution over likely adversaries and (ii) its current mitigation maturity state, derived from CSF assessments and mapped to ATT\&CK mitigations as described in §\ref{subsec-csf-mapping}. Actions correspond to planned mitigations that harden systems and reduce future attack feasibility. Real-time detection, containment, and incident response are outside the scope of this environment.

\subsection{Markov decision process definition}
Using the environment defined in Section \ref{sec:env-setup}, we induce an MDP in which the defender selects a mitigation portfolio subject to maturity and budget-constrained feasibility.

\paragraph{Episode definition}
Each training episode simulates adversarial campaigns against a single defender. 
At the beginning of an episode, the defender observes its current organizational maturity and the set of likely adversaries, then selects a mitigation technique set $\mathbf{a}$ subject to budget constraints. This action remains fixed for the duration of the episode. The population of adversaries independently executes their attack flows. Adversary technique proposals are sampled from the sequencing model in Section \ref{subsec-vomm}. The episode proceeds over multiple time steps as adversaries consume their resources while attempting successive techniques. The defender wins if all adversaries exhaust their resources before causing an impact, and loses if any adversary reaches an impact technique state.

\paragraph{Observation definition}
The observation (i.e., what a defender sees) at time $t$ is a tuple
\begin{equation}
\label{eq:observation_definition}
o_t
=
\big(\,\mathbf{m},\, \mathbf{Z}\,\big)
\end{equation}

where:
\begin{itemize}
  \item $\mathbf{m}\in[0,1]^{M}$ is the defender’s \emph{mitigation maturity vector} over the $M$ ATT\&CK mitigations (component $m_i$ encodes readiness/effectiveness for mitigation $i$).
  \item $\mathbf{Z}$ is a stacking binary indicator of adversary TTPs: for each of the $N$ adversaries in the episode, we include a binary vector indicating which techniques are observed in the profile of that adversary.
\end{itemize}
This design captures what the defender can plausibly know about who is likely to target the organization (i.e., top-$N$ adversaries, Section \ref{subsec:adversary-modeling}) and which techniques those adversaries tend to employ in addition to the organization’s own CSF maturity levels.

\paragraph{Action definition}
The action is a multi-hot vector $\mathbf{a}\in\{0,1\}^{M}$ selecting a set of MITRE ATT\&CK mitigations to apply at the start of the episode. Defender actions are constrained by a budget mechanism described below. The model may determine to allocate effort to a few high-impact mitigations or several smaller ones.

Each mitigation has a (cost, complexity) pair. We map these ordinal ratings to a fractional share of the episode budget via a lookup table $\mathrm{PctCost}(\cdot)\in(0,1]$. For interpretability, we express costs in percentage units:
\begin{equation}
\label{eq:mitigation_cost}
\mathrm{Cost}_i
=
100 \cdot
\mathrm{PctCost}(\text{cost}_i,\text{complexity}_i)
\cdot
\mu(m_i)
\end{equation}

The episode budget is normalized to 100 percentage units, and a mitigation set is feasible if the total cost does not exceed this budget.
\begin{equation}
\label{eq:budget_constraint}
\sum_{i:\,a_i=1} \mathrm{Cost}_i
\;\le\;
100
\end{equation}

At the start of each episode, the defender has the full budget available, and each selected mitigation deducts a fraction of the budget proportional to its cost. The maturity scaling function $\mu(\cdot)$ is defined by linear interpolation between anchor points chosen so that a benchmark mitigation consumes the full episode budget at a given maturity level.

As an example, for Maturity~4 we define the anchor as a mitigation with Very High cost and Very High complexity, reflecting the largest action a highly mature organization could reasonably complete within a single planning episode. The corresponding value of $\mu(4)$ is then chosen so that this anchor mitigation exactly exhausts the episode budget. Analogous anchors are defined for lower maturity levels, yielding a piecewise-linear scaling function that governs how mitigation costs vary with maturity.

Ordinal cost and complexity ratings are mapped to base budget fractions using a fixed
lookup table (Figure~\ref{fig:basepct} in Appendix~\ref{app:budget-map}).

In our experiments, the full-budget anchors are:
\begin{itemize}
  \item Maturity~1: Medium/Medium mitigation, resulting in $\mu(1)=5.0$,
  \item Maturity~2: Medium/High mitigation, resulting in $\mu(2)=3.6$,
  \item Maturity~3: High/High mitigation, resulting in $\mu(3)=2.8$,
  \item Maturity~4: Very High/Very High mitigation, resulting in $\mu(4)=2.0$.
\end{itemize}
Thus, lower-maturity defenders can still combine several smaller actions as long as their \emph{sum} stays $\le 100\%$, while actions larger than the benchmark at a given maturity are effectively out of scope for that episode.

As an example, a control at \emph{Very Low} cost and \emph{Low} complexity consumes $8\%$ of the budget. If the organization's maturity is a continuous score $m\!\in[0,1]$, the effective multiplier is
\begin{equation}
\label{eq:linear_budget_multiplier}
\mu(m)
=
y_0
+
\frac{m - x_0}{x_1 - x_0}
\bigl(y_1 - y_0\bigr)
\end{equation}

where $(x_0,y_0)$ and $(x_1,y_1)$ are the two adjacent anchor points that bracket $m$. Thus, the budget consumption rate is
\[
8\% \times \mu(m).
\]

If $m=0.65$ lies between anchors $(0.6,\,\mu_{0.6})$ and $(0.7,\,\mu_{0.7})$, then
\[
\mu(0.65)=\mu_{0.6}+\frac{0.65-0.6}{0.7-0.6}\,\bigl(\mu_{0.7}-\mu_{0.6}\bigr)
=\tfrac{1}{2}\,(\mu_{0.6}+\mu_{0.7}),
\]
so the scaled cost is $8\% \times \mu(0.65)$.

In the step function, we also calculate the adversary per-target budget. Each adversary is characterized by a \textit{type}, a \textit{resource level} (e.g., approximate staffing), and a \textit{sophistication} level. We estimate the adversary’s expected per-target resource availability usin a resource-spread factor $\mathrm{Spread}(\text{type},\text{resource})$, which captures how the adversary’s resources are distributed across concurrent targets (Appendix~\ref{app:resource-spread}). For example, an adversary with roughly 10 operators targeting roughly 1000 organizations per month results in an average of $0.01$ operator-equivalents per target per month.

Technique costs to the adversary are computed using an attacker lookup table $\mathrm{PctCost}_{\text{adv}}(\text{tech})$, distinct from the defender’s mitigation costs. We scale attacker costs by a sophistication multiplier $\sigma$ that models operational efficiency such that low-sophistication actors incur higher effective costs, whereas high-sophistication actors incur lower costs. We use the same anchor values as the defender’s budget scaling for interpretability, with $\sigma \in \{5.0, 3.6, 2.0\}$
for \{Low, Medium, High\} sophistication. The resulting effective per-technique budget consumption is
\begin{align}
\label{eq:adversary_cost_budget}
\mathrm{Cost}_{\text{adv}}(\text{tech})
&=
100 \cdot
\mathrm{PctCost}_{\text{adv}}(\text{tech})
\cdot
\sigma \\
\mathrm{Budget}_{\text{adv}}
&=
100 \cdot
\mathrm{Spread}(\text{type}, \text{resource})
\end{align}

An adversary advances on a technique only if it has sufficient residual per-target budget. Otherwise, it stalls for that step.

At each environment step, given the defender’s selected mitigation set $\mathbf{a}$ (chosen at episode start), each adversary attempts to execute its next technique. We model attack flows as sequential and independent, with no coordination among adversaries. A probabilistic effectiveness model reduces technique success probability when the technique is covered by any selected mitigation. Mitigations of higher effectiveness reduce the probability of attack success. If the technique succeeds, the adversary’s residual per-target budget is decremented by $\mathrm{Cost}_{\text{adv}}(\text{tech})$ and it advances. Otherwise, the adversary is blocked for that step.

\paragraph{Step function}
Adversary technique proposals follow the sequencing model in Section \ref{subsec-vomm}, while success probabilities are modulated by the defender’s selected mitigations and their maturities.

At each step $t$:
\begin{enumerate}
  \item For each adversary $j$, sample the next technique $x_t^{(j)}$ from its VOMM prior $p_{\text{VOMM}}(x_t^{(j)} \mid \mathbf{h}_{t-1}^{(j)})$, where $\mathbf{h}_{t-1}^{(j)}$ is its recent technique history.
  \item Compute the success probability for $x_t^{(j)}$ as
  \[
  P_{\text{succ}}^{(j)} = 1 - \mathrm{EffCov}\!\left(x_t^{(j)}, \mathbf{a}\right),
  \]
  where $\mathrm{EffCov}(\cdot)$ measures mitigation coverage and effectiveness.
  \item If the attack succeeds, decrement the adversary’s budget by its technique cost and advance to the next state. If it fails, mark the technique as blocked.
  \item The environment emits a reward $r_t$ reflecting blocking success and mitigation relevance (Section~\ref{subsec:reward}).
\end{enumerate}
The episode terminates if any adversary causes impact (\emph{loss}) or all adversaries deplete their budgets (\emph{win}).

\subsection{Reward and cost modeling}
\label{subsec:reward}
The reward function encourages (i) blocking adversary progress, (ii) selecting mitigations that are relevant and effective against techniques actually attempted during the episode, and (iii) successfully exhausting all adversaries’ budgets. Let $S_t$ denote the vector of per-adversary success indicators at time $t$
($1$ if the adversary is blocked at that step, $0$ otherwise), and let
$\|\mathbf{a}\|_1$ be the number of selected mitigations. The environment returns
\begin{equation}
\label{eq:reward_function}
\begin{aligned}
r_t
&= 100 \cdot \sum S_t \\
&\quad + \frac{
\mathrm{CoverEff}(\mathbf{a}, \text{active techniques})
}{
\|\mathbf{a}\|_1 + 1
} \\
&\quad + 1000 \cdot \mathbb{I}\{\text{win at } t\}
\end{aligned}
\end{equation}

The terminal bonus incentivizes depleting all adversaries’ budgets without impact, while the first term rewards per-step blocking across all adversaries. The second term rewards mitigation relevance and effectiveness while regularizing against indiscriminate selection of many mitigations via the denominator $\|\mathbf{a}\|_1 + 1$.

The coverage effectiveness score $\mathrm{CoverEff}(\cdot)$ rewards selecting mitigations that both cover adversary techniques and are effective against the specific threats observed during the episode. For each attack technique covered by at least one selected mitigation, the score is computed as follows. If the technique was not attempted by any adversary during the episode, a small baseline reward of $1$ is added. If the technique was attempted, we instead add
\[
(\max \mathrm{Eff} + 1) \times 5,
\]
where $\max \mathrm{Eff}$ is the highest effectiveness value among all selected
mitigations that cover the technique.

Mitigation effectiveness is computed using a deterministic matching hierarchy that prioritizes specificity. We first use effectiveness scores for exact adversary–technique matches when available. If no exact match exists, we fall back to matches based on adversary type and sophistication. If neither is available, we use the global average effectiveness for that technique across all adversaries in our threat intelligence database. When multiple mitigations cover the same technique, only the most effective mitigation contributes to the score.

The defender’s policy therefore learns to allocate a finite mitigation budget to actions that (i) most effectively reduce the probability of adversary technique success, given the observed adversary TTP profiles, and (ii) drain adversary budgets before impact. The number of adversaries per episode is treated as an input derived from upstream threat intelligence and reflect the most likely concurrent threats targeting the organization. This design is consistent with empirical evidence that adversary groups exhibit persistent targeting patterns and that a small number of probable threats dominate organizational risk. In this work, we fix the adversary set per episode to ten.

\subsection{Deep reinforcement learning agent}
We instantiate the defender policy using a Deep Q-Network (DQN), which is a value-based reinforcement learning method that estimates the expected utility of selecting each mitigation given the current organizational state. While the nominal action space corresponds to the power set of mitigations, our formulation significantly reduces this complexity. First, actions are selected once per episode as a constrained portfolio rather than sequential combinatorial decisions. Second, budget constraints and costs prune infeasible actions, which restricts the effective decision space to a small subset of high-value mitigations. Third, the Q-network outputs each mitigation values to provide an efficient greedy selection without enumerating all possible subsets. This structure aligns well with DQN, which scales effectively when actions can be decomposed into independent value estimates. Compared to policy-gradient methods designed for large continuous or unconstrained combinatorial spaces, DQN provides stable training, sample efficiency, and direct interpretability of mitigation selection.

With DQN, we approximate $Q(o_t, a)$, where the observation $o_t = (\mathbf{m}, \mathbf{Z})$ consists of the defender’s mitigation maturity vector $\mathbf{m}$ and the stacked binary technique indicators $\mathbf{Z}$. These components are concatenated into a single feature vector, 
passed through fully connected layers, and mapped to a linear output of size $M$, corresponding to the Q-value of selecting each mitigation.

The agent greedily selects mitigations in descending $Q$ subject to the defender budget, with $\epsilon$-greedy exploration. The agent uses experience replay to stabilize training, with a discount factor of $\gamma = 0.90$ (favoring near-term rewards), a learning rate of $10^{-3}$, and an exploration rate $\epsilon$ that gradually decreases from $1.0$ (fully exploratory) to $0.05$ (mostly exploitative) over time.

At each simulation step, each adversary advances its attack by proposing a next technique according to a VOMM learned from real-world attack flows. Concretely, given its recent history $\mathbf{h}$ (last $k \le K$ techniques), the adversary samples the next technique $x$ from $p_{\text{VOMM}}(x \mid \mathbf{h})$, which produces realistic sequences without enumerating the full ATT\&CK action space at every step.

We consider two standard policy formulations that combine the learned preferences $Q(s,a)$ with the prior $p_{\text{VOMM}}(a \mid \mathbf{h})$ derived from recent history $\mathbf{h}$ (last $k \le K$ moves). 

The first form is an additive mixture:
\begin{equation}
\label{eq:hybrid_policy}
\pi(a \mid s,\mathbf{h})
=
(1-\lambda)\,
\frac{
\exp\!\left(\tfrac{Q(s,a)}{\tau}\right)
}{
\sum_{a'} \exp\!\left(\tfrac{Q(s,a')}{\tau}\right)
}
+
\lambda\, p_{\text{VOMM}}(a \mid \mathbf{h})
\end{equation}

where $\lambda \in [0,1]$ controls the weight between learned and prior components and $\tau$ is the softmax temperature.

Alternatively, we use a multiplicative (product-of-experts) formulation:
\begin{equation}
\label{eq:product_hybrid_policy}
\pi(a \mid s,\mathbf{h})
\propto
\exp\!\left(\tfrac{Q(s,a)}{\tau}\right)\,
p_{\text{VOMM}}(a \mid \mathbf{h})^{\beta}
\end{equation}

where $\beta$ controls the relative sharpness of the VOMM prior.

During early training, we set a high mixing weight ($\lambda$ or $\beta$) so that the policy leans heavily on the VOMM prior, which encourages realistic adversary behavior and structured exploration. As experience accumulates, this weight gradually decreases to allow the learned $Q$-values to increasingly dominate and drive performance toward optimal rather than imitative behavior.

\subsection{Attack Path Reconstruction}
\label{sec:attack_path_reconstruction}

Once trained, the DQN can recommend a mitigation portfolio for a defender given an observation that combines (i) the top adversaries targeting that defender and their observed ATT\&CK techniques and (ii) the defender’s assessed NIST CSF practice maturity levels. However, this mapping from state to a generic multi-label mitigation action is not, by itself, sufficient to support operational decision-making. First, ATT\&CK mitigations such as \texttt{M1017: User Training} are abstract and can be implemented in many ways with very different costs and effects. Second, the defender must understand why a particular mitigation is recommended, i.e., which likely attack paths it is intended to disrupt. Thus, our goal at this stage is to ground each mitigation in concrete adversary behavior by tying it to specific techniques previously observed for that adversary. Moreover, we reconstruct likely attack and defense paths that make the defender’s mitigation effect more interpretable.

In this phase, reconstruction is conditioned on an up-front defender mitigation selection and then uses beam search to simulate likely adversary progressions against that selected defense posture. The reconstruction process integrates (i) defender utility estimates from the trained DQN, (ii) probabilistic adversary transition modeling through the VOMM and prior observed attack behavior, and (iii) the stateful cyber defense simulation environment that evaluates candidate actions.

Unlike traditional graph expansion approaches, reconstruction operates directly over the simulation environment to enable consistent handling of adversary progression, defender actions, and resource constraints. Each node represents a complete snapshot of the environment state, including adversary histories, remaining budgets, and mitigation maturity.

\paragraph{Up-front Mitigation Selection.}
For a given root observation comprising the organization's mitigation maturity levels, as described in Section~\ref{subsec-csf-mapping}, and the set of likely adversaries with their observed techniques, as described in Section~\ref{subsec:adversary-modeling}, we first query the trained DQN once to obtain mitigation Q-values $Q(o_0,\cdot)$. These Q-values are sorted to construct a fixed root mitigation portfolio. This portfolio represents the initial set of mitigations that the defender can realistically commit to under the available budget.

During reconstruction, the beam search does not enumerate arbitrary new mitigation portfolios at every node. Instead, for each candidate adversary technique, it selects from this fixed portfolio the mitigation that is most applicable to the technique. Preference is given to mitigations associated with observed adversary techniques. If no such exact match exists, the search falls back to other mitigations in the root portfolio that cover the technique. This design preserves tractability while ensuring that reconstructed defenses remain tied to the DQN's learned priorities and to specific observed adversary behavior.

\paragraph{State Representation.}
Each beam node stores a full environment state obtained via a snapshot mechanism. The state includes:
(i) mitigation maturity levels,
(ii) adversary technique histories,
(iii) adversary resource budgets and phase progression, and
(iv) latent variables governing adversary behavior (e.g., next technique selection).
This design allows exact restoration of prior states and ensures that candidate expansions are evaluated without side effects.

\paragraph{Candidate Expansion via Simulation.}
From each beam node, candidate actions are evaluated by simulating a single step of the environment. For a given node state $s$, the algorithm:
\begin{enumerate}
    \item Restores the environment to state $s$,
    \item Constructs a candidate defender action $a$ using DQN-guided selection under budget constraints,
    \item Advances the adversary by one step using its probabilistic policy,
    \item Observes the resulting state $s'$, reward $r$, selected adversary technique, and termination condition.
\end{enumerate}

This simulation is performed using a side-effect-free mechanism that restores the original state after each candidate evaluation. As a result, multiple candidate expansions can be explored from the same parent node without interference.

\paragraph{Adversary Modeling with Observed Technique Bias.}
Adversary behavior is governed by a probabilistic policy derived from the VOMM described in Appendix \ref{subsec-vomm}.

This distribution is augmented to favor techniques previously observed for the adversary. Candidate techniques are ranked with priority given to higher transition probability and the technique's membership in the set of attack techniques previously observed by the given adversary.

\paragraph{Mitigation Selection with Technique-Specific Effectiveness.}
Defender actions are evaluated at each step by selecting mitigations that directly apply to the adversary’s chosen technique. For a given technique, the reconstruction evaluates all selected mitigations and identifies the mitigation with the highest effectiveness against the specific adversary–technique pair.

Effectiveness is derived from a hierarchical mapping that prioritizes:
(i) exact adversary–technique mitigation relationships,
(ii) adversary-type and sophistication matches,
(iii) aggregated effectiveness when specific mappings are unavailable.

A mitigation is considered successful if its effectiveness exceeds a reconstruction threshold, ensuring that reconstruction emphasizes concrete, high-confidence defensive effects rather than abstract coverage.

\paragraph{Defender Scoring via DQN.}
For each candidate node, defender utility is estimated using a trained Deep Q-Network. The DQN produces Q-values over mitigation actions given the current observation (mitigation maturity and adversary technique profile). These Q-values are used to construct feasible mitigation sets (e.g., via greedy or knapsack-style selection) and serve as a scoring signal that biases the beam search toward states with higher expected defensive value.

\paragraph{Beam Search Algorithm.}
Attack-path reconstruction is performed using a beam search of width $k$ over depth $d$, where each node represents a full environment state capturing both adversary dynamics and defender effects over time. The procedure initializes a root mitigation portfolio from the DQN Q-values and iteratively expands candidate states by sampling likely adversary techniques using the VOMM with observed-technique bias. For each candidate technique, the best corresponding mitigation from the root portfolio is selected and applied via simulation to produce successor states. Each state is scored using a combination of cumulative reward, likelihood, uncertainty, and impact terms with a diversity penalty, and the top-$k$ states are retained at each depth. A state is terminal if the adversary reaches an impact objective, exhausts its resource budget, or the maximum search depth is reached. Running this procedure independently for each prioritized adversary yields a set of high-scoring attack paths that are both consistent with historical technique usage and aligned with the learned mitigation policy. Aggregating across these paths produces a finite set of candidate mitigations annotated with the adversaries and techniques they counter, their estimated impact, and their contribution to cumulative reward (see Appendix~\ref{app:beam_reconstruction}).

\subsection{Mitigation Plan Optimization.}
In practice, defenders are constrained by a finite budget and cannot deploy every candidate mitigation surfaced by the beam search. We therefore treat the union of mitigations appearing on the top $k$ paths as a candidate pool for downstream optimization. The resulting candidate mitigation set defines the feasible action space for optimization.

Mitigations selected along individual attack paths are conditioned on a specific adversary and evaluated under a relatively permissive defender budget during reconstruction. The union of these mitigations across paths, however, does not directly translate into an actionable defense plan. In practice, organizations execute mitigation work in discrete planning cycles under stricter resource constraints, which require a prioritization across all candidate mitigations induced by the reconstructed attack paths.

We therefore formulate mitigation planning as a combinatorial optimization problem with budget constraints. Let $\mathcal{M} = \{1,\dots,N\}$ denote the set of candidate mitigations extracted from beam search across all adversaries. Each candidate $i \in \mathcal{M}$ is associated with a cost $c_i \in \mathbb{R}^+$, representing implementation effort given organizational maturity, and a value $v_i \in \mathbb{R}^+$, representing its expected contribution to reducing adversary success.

The value $v_i$ aggregates signals from attack-path reconstruction and captures three factors: (i) the likelihood that the mitigation remediates the associated technique, (ii) the degree of support from high-scoring reconstructed paths, and (iii) the frequency with which the mitigation appears across these paths. Concretely, we compute $v_i$ as a weighted combination of remediation likelihood, normalized path score contribution, and a logarithmic occurrence term to favor mitigations that are both effective and broadly applicable to multiple adversarial techniques across the attack paths.

The optimization objective selects a subset of mitigations that maximizes total defensive value subject to a finite budget $B$:
\begin{equation}
\max_{x_i \in \{0,1\}} \sum_{i \in \mathcal{M}} v_i x_i
\quad \text{s.t.} \quad
\sum_{i \in \mathcal{M}} c_i x_i \leq B.
\end{equation}

We solve this as a 0--1 knapsack problem to produce a deterministic mitigation portfolio within a given budget. The resulting mitigation portfolio maximizes expected defensive impact within a fixed work-cycle budget. Each selected mitigation remains traceable to specific reconstructed attack paths and adversary behaviors that provide both prioritization and explainability for deployment.

\subsection{Analysis}
To evaluate the decision quality of our DRL defender, we compare it against a strong optimization baseline, referred to as an oracle policy. The oracle is not intended to represent a globally optimal defender, but rather, it provides a non-learning reference policy that selects mitigations using explicit domain knowledge and constrained optimization. This baseline enables objective comparison between learned and non-learned mitigation strategies under identical environmental conditions.

\paragraph{Oracle Policy.}
For a given evaluation episode, the oracle operates with respect to a fixed set of adversaries. It begins by extracting each adversary’s observed set of attack techniques and identifying all mitigations associated with those techniques via MITRE ATT\&CK mappings. Candidate mitigations are filtered based on organizational feasibility, including maturity constraints and budget availability.

Given the resulting candidate set, the oracle selects mitigations by solving a budget-constrained optimization problem that maximizes a deterministic proxy objective capturing expected mitigation effectiveness. Specifically, each candidate mitigation $m$ is assigned a proxy benefit defined as
\begin{equation}
\label{eq:mitigation_benefit}
\text{benefit}(m)
=
\sum_{a \in \mathcal{A}}
\sum_{t \in T_a}
w(a,t)
\cdot
p(a,t,m)
\end{equation}

where $\mathcal{A}$ denotes the set of adversaries considered in the episode, $T_a$ is the set of techniques observed for adversary $a$, $w(a,t)$ is an importance weight reflecting the expected relevance of technique $t$ for adversary $a$, and $p(a,t,m)$ is the deterministic likelihood that mitigation $m$ remediates technique $t$ when employed against adversary $a$.

The importance weight $w(a,t)$ is derived from the adversary behavior model and incorporates the VOMM-likelihood of technique usage that is reweighted by an adversary-specific prior $P(t \mid a)$ computed from historical technique observations. The term mitigation effectiveness $p(a,t,m)$ corresponds to the likelihood of deterministic remediation used by the simulation environment, which is derived from the adversary technique and mitigation relationships and the sophistication of the adversary.

The oracle then selects a subset of mitigations that maximizes the aggregate proxy benefit subject to budget and feasibility constraints. This optimization is performed using an integer linear program (ILP), producing a single mitigation set for the episode.

\paragraph{Evaluation Methodology.}
The oracle policy and the learned DRL policy are both evaluated using the same expected return function $J(\pi)$. Evaluation is performed by executing each policy over multiple independent simulation episodes spanning synthetically generated organizational profiles and real-world adversary configurations.

Adversary behavior within the simulator is generated by a VOMM-based adversary policy, and we additionally evaluate robustness under adversary model shift by varying adversary behavior distributions at test time. Although the oracle selects mitigations deterministically using a non-stochastic proxy objective, all policies interact with the same stochastic environment dynamics during evaluation, including the probabilistic mitigation success.

The evaluation function is defined as
\begin{equation}
\label{eq:evaluation}
J(\pi)
=
\mathbb{E}\left[
-\alpha \cdot \text{Loss}
-\beta \cdot \text{Cost}
\right]
\end{equation}
where $\text{Loss}$ captures realized attack impact (e.g., attacker success or critical asset compromise), $\text{Cost}$ represents the total mitigation expenditure incurred during the episode, and $\alpha,\beta$ are weighting coefficients. We estimate the expected return via repeated simulation.

In practice, we estimate the expected return using Monte Carlo simulation,
\begin{equation}
\label{eq:policy_evaluation}
\hat{J}(\pi)
=
\frac{1}{N}
\sum_{i=1}^{N}
\left(
-\alpha \cdot \text{Loss}^{(i)}
-
\beta \cdot \text{Cost}^{(i)}
\right)
\end{equation}

where $N$ denotes the number of independent simulation episodes.

All reported results use identical simulation conditions across policies, which enable direct comparison of mitigation decision quality and cost-effectiveness.

\section{Experimental Evaluation}
\label{sec:experiment}

\subsection{Experimental setup and base configuration}
\label{subsec:exp-setup}
All results are reported relative to a fixed base configuration that anchors all comparisons. Each episode simulates a single defender facing ten adversaries, with defender and adversary budgets set to $100$ units. Episodes terminate when the adversary reaches an impact state (defender loss) or exhausts its per-episode budget (defender win). Consistent with our decision formulation, the defender selects a single mitigation portfolio at episode start (shields up), and that action is held constant for the remainder of the episode.

We evaluate the DQN and oracle using identical episode instances by resetting the environment once, snapshotting the initial state, and running both policies from that same state. This ensures that both policies face the same adversary and organizational maturity profile, and the differences in evaluation are due solely to policy decisions.

Training uses the DQN described in Section~\ref{sec:decision-formulation} with replay buffer capacity $10{,}000$, learning rate $10^{-4}$, batch size $64$, warm-up of $500$ steps, and $\epsilon$-greedy exploration initialized at $0.99$ with linear decay over $500$ steps. During evaluation, exploration is disabled ($\epsilon=0$). Defender maturity profiles are sampled from the synthetic organization generator (Section~\ref{subsec:synthetic-generation}), and adversary sequencing is governed by the VOMM (Section~\ref{subsec-vomm}).

Performance is reported as the empirical mean return over 1,000 independently sampled episodes, with $\alpha=1.0$ and $\beta=0.01$.

\subsection{Base policy performance}
\label{subsec:base-performance}
The base policy corresponds to the default evaluation configuration (defender and adversary budgets of 100, greedy action selection, warm-up of 500, decay steps of $1{,}000$, batch size of 64, and the reward defined in Section~\ref{subsec:reward}).

On the $1{,}000$ paired evaluation episodes, the base DQN achieves a win rate of $37.5\%$ (oracle: $33.9\%$) with comparable episode lengths, while selecting a similar number of mitigations on average. In terms of the evaluation objective, the base DQN attains $J(\pi)=-0.6309$ versus $J(\pi)=-0.6677$ for the oracle, yielding regret $J(\text{oracle})-J(\text{DQN})=-0.0369$.

\subsection{Ablation study}
\label{subsec:ablation}
We perform single ablations with all non-ablated settings fixed to the base configuration. To ensure fair comparisons across model variants, we evaluate each policy on the same static episode set by restoring the simulator state from a pre-generated episode corpus. We report the expected return $J(\pi)$ and regret relative to the oracle, along with win/loss rates, mitigation cost, and episode length. The full results are shown in Table~\ref{tab:ablation-results}.

Overall, the ablation results reveal a markedly different operating regime compared to the base configuration. Across all variants, win rates are substantially lower and episode lengths are longer, indicating that the evaluation setting induces more persistent adversary progression and places greater emphasis on sustained defensive effectiveness rather than early disruption. In this regime, differences between configurations are primarily reflected in cost versus performance trade-offs captured by $J(\pi)$.

\paragraph{Budget sensitivity.}
Adversary budget remains the dominant factor influencing performance. Reducing the adversary budget (Adv.\ Budget 50) yields the best-performing policy with $J(\pi)=-0.9613$, outperforming the oracle by $0.0285$. This improvement is accompanied by longer episodes and higher mitigation costs, suggesting that the learned policy allocates resources to sustain defensive pressure over extended attack sequences. In contrast, increasing adversary budget (Adv.\ Budget 150) significantly degrades performance, as higher-capability adversaries are more likely to achieve impact.

The defender budget exhibits a similar but less pronounced effect. Increasing defender budget (Def.\ Budget 150) improves return relative to the base model, while reducing it (Def.\ Budget 50) leads to degraded performance. However, the magnitude of these changes is smaller than those observed under adversary budget variation, reinforcing that adversary capability is the primary driver of task difficulty.

\paragraph{Learning dynamics.}
Training-related parameters show consistent but moderate effects. The base model outperforms both smaller and larger batch sizes, with Batch Size 32 and 128 yielding lower returns. Similarly, simplified reward shaping reduces performance, indicating that the full reward formulation provides a more informative signal for distinguishing effective mitigation strategies.

\paragraph{Environment scale.}
Varying the number of organizations used during training (Org.\ Count 50 vs.\ 100) results in only minor differences in performance. This suggests that the learned policy generalizes across moderate variations in training diversity, and that the base configuration provides sufficient coverage of organizational variability.

Overall, the ablation study demonstrates that policy performance is most sensitive to adversary capability, with secondary effects from defender budget and learning dynamics. The learned policy consistently outperforms the oracle in terms of expected return, indicating that sequential decision-making under uncertainty provides a measurable advantage over static mitigation selection strategies in complex adversarial environments.

\begin{table*}[t]
\centering
\caption{Ablation study results over $N=7000$ evaluation episodes per policy variant.}
\label{tab:ablation-results}
\small
\renewcommand{\arraystretch}{1.12}
\setlength{\tabcolsep}{0pt}

\begin{tabular*}{\textwidth}{@{\extracolsep{\fill}}lrrrrrrrr@{}}
\hline
\textbf{Model} &
\textbf{Win} &
\textbf{Loss} &
\textbf{Cost} &
\textbf{Cost \%} &
\textbf{Avg. Mit.} &
\textbf{Path Len.} &
$\mathbf{J(\pi)}$ &
\textbf{Regret} \\
\hline
Adv. Budget 50        & 4.7 & 95.3 & 85.91  & 85.9  & 1.6 & 21.6 & -0.9613 & -0.0285 \\
Def. Budget 150       & 4.8 & 95.2 & 121.99 & 122.0 & 2.6 & 21.2 & -0.9645 & -0.0253 \\
Base Model            & 2.6 & 97.4 & 75.47  & 75.5  & 1.8 & 18.5 & -0.9811 & -0.0086 \\
Batch Size 32         & 1.8 & 98.2 & 73.58  & 73.6  & 2.0 & 17.0 & -0.9889 & -0.0008 \\
Oracle                & 1.8 & 98.2 & 79.08  & 79.1  & 1.1 & 15.9 & -0.9898 &  0.0000 \\
Simple Reward Shaping & 1.1 & 98.9 & 64.14  & 64.1  & 2.1 & 15.4 & -0.9950 &  0.0052 \\
Adv. Budget 150       & 0.7 & 99.3 & 39.90  & 39.9  & 0.7 & 14.3 & -0.9973 &  0.0075 \\
Batch Size 128        & 0.6 & 99.4 & 39.90  & 39.9  & 0.7 & 14.2 & -0.9977 &  0.0079 \\
Def. Budget 50        & 0.4 & 99.6 & 25.34  & 25.3  & 1.3 & 14.1 & -0.9984 &  0.0086 \\
Org. Count 50         & 0.6 & 99.4 & 51.15  & 51.1  & 1.5 & 14.1 & -0.9993 &  0.0095 \\
Org. Count 100        & 0.7 & 99.3 & 71.81  & 71.8  & 1.6 & 15.0 & -0.9998 &  0.0100 \\
\hline
\end{tabular*}
\end{table*}

\begin{figure}[t]
  \centering
  \includegraphics[width=\linewidth]{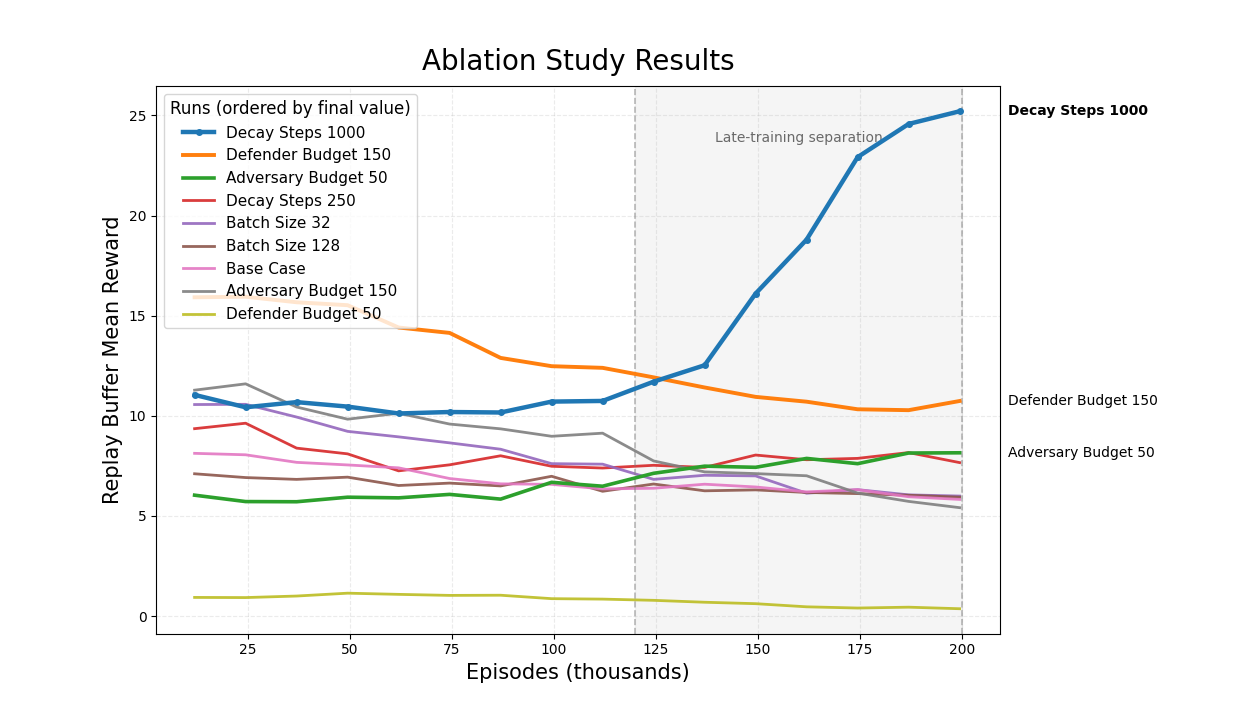}
  \caption{
  Ablation study results showing learning dynamics across configuration settings, measured by the mean reward in the replay buffer over training episodes.}
  \label{fig:ablation_learning}
\end{figure}

To further examine learning dynamics under increased adversarial pressure, we conducted an extended training run with a larger number of concurrent adversaries and an additional ablation on the number of decay steps. Figure~\ref{fig:ablation_learning} shows that, although absolute reward levels decrease in this more challenging setting, the relative ordering of configurations remains largely consistent with the primary ablation results. In particular, a clear late separation of training emerges, which indicates that longer exploration schedules improve policy quality over time. In particular, increasing the decay steps to $1{,}000$ shows the best performance, suggesting that sustained exploration is critical for learning effective mitigation strategies in more complex multi-adversary environments. These results indicate that the observed performance trends are robust to the increased complexity of the environment.

\subsection{Case Study}
We evaluated the proposed framework on two representative U.S. school districts with comparable size and threat profile but differing cybersecurity practice maturity distributions. Each organization was assessed using NIST CSF 2.0 practices, which were translated into mitigation capability profiles and used to drive adversary simulation and mitigation planning.

Despite similar organizational attributes, the resulting mitigation portfolios differed slightly. School 1 received a mitigation plan with four techniques (M1030: Network Segmentation, M1040: Endpoint Behavior Prevention, M1017: User Training, and M1024: Restrict Registry Permissions), while School 2 received an additional mitigation (M1036: Account Use Policies). This difference is attributable to variations in practice maturity, particularly in detection and response capabilities, which reduced baseline effectiveness against certain adversary techniques in School 2.

Attack path reconstruction reveals that both organizations face similar dominant adversary behaviors, including credential abuse and process injection following initial reconnaissance. For example, a high-likelihood path includes network sniffing (T1040), process injection (T1055), and abuse of valid accounts (T1078), with mitigations such as network segmentation (M1030), endpoint behavior prevention (M1040), and user training (M1017) directly reducing attack success probabilities along this sequence.

Moreover, the mitigation recommendations are tied to specific adversary techniques and provide actionable guidance to counter the threat. For instance, process injection (T1055) is mitigated through endpoint behavior prevention (M1040), which detects anomalous memory access patterns and blocks unauthorized code execution within legitimate processes. Similarly, abuse of valid accounts (T1078) is mitigated through user training (M1017), reducing susceptibility to credential compromise and social engineering.

\begin{figure}[t]
  \centering
  \includegraphics[width=\linewidth]{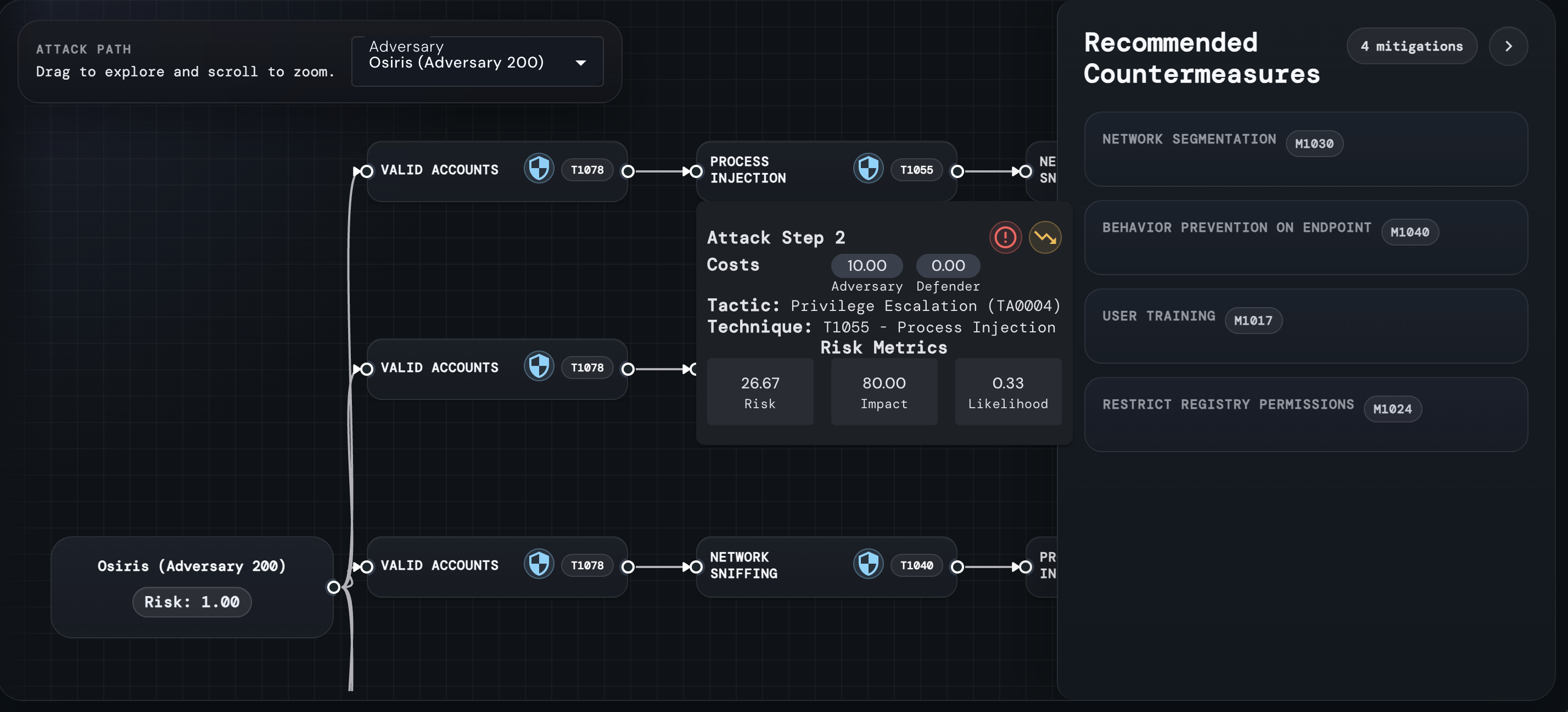}
  \caption{%
  Attack-path reconstruction for School 1 with multiple paths output from the beam search, and the corresponding mitigation portfolio for a school district. The system identifies high-likelihood adversary paths and recommends a budget-constrained set of mitigations that disrupt key techniques.
  }
  \label{fig:attack-path}
\end{figure}

A representative beam search path with associated attributes and mitigation selections for School 1 is shown in Figure~\ref{fig:attack-path}.

Overall, the case study demonstrates that the framework produces consistent core mitigation strategies across similar environments while adapting to subtle differences in organizational maturity. This supports the claim that operationalizing governance assessments enables targeted, budget-aware mitigation planning grounded in adversary behavior and system-specific risk.

\section{Conclusion and Future Work}
This work addresses a practical gap between cybersecurity governance frameworks and daily defensive planning. Defenders must translate high-level assessments and threat intelligence into a budget-feasible set of concrete mitigation work items for a planning period. We introduced a strategic mitigation planning formulation in which defender actions are mitigation portfolios selected under explicit cost constraints, and we coupled this formulation to a simulation environment that reflects (i) heterogeneous organizational capability derived from NIST CSF practice maturity, (ii) empirically grounded adversary behavior using ATT\&CK technique sequences, and (iii) mitigation effectiveness signals linked to adversary tradecraft.

Our approach makes three core contributions. First, we define a constrained MDP for mitigation planning in which actions correspond to mitigation selections subject to an explicit budget model, and align the learning problem with how organizations plan defensive work in sprints or cycles. Second, we introduced an LLM semantic translation layer that maps assessable NIST CSF practices to ATT\&CK mitigations, allowing commonly collected CSF maturity data to parameterize mitigation feasibility and cost in the learning environment. Third, we incorporated a technique-sequencing prior learned from ATT\&CK Flow data via a variable-order Markov model, which constrains simulated adversary progression to realistic paths and improves sample efficiency while preserving exploration. After training, we further improve operational usability by reconstructing high-likelihood attack/defense paths and surfacing traceable mitigation recommendations that connect suggested work items to the specific adversary techniques they are intended to disrupt.

Our experimental results evaluated against an oracle baseline suggest that a learned DQN policy can match or exceed a strong optimization reference under identical episodic conditions, while also producing actionable mitigation sets that respect budget constraints. Overall, the framework supports a workflow in which RL is used to learn cost-effective mitigation preferences offline, and human defenders validate and implement a transparent and highly contextual mitigation portfolio online.

Future work includes extending the framework to more dynamic mitigation deployment over time, training adversarial behavior on more robust system models, and exploring alternative policy representations for larger or more continuous action spaces.

%% file: sections/appendices.tex
\section{Synthetic Data Generation Framework}
\label{app:synthetic_data}

\subsection{Synthetic Data Model}

\paragraph{Organization latent maturity.}
Each organization $i$ is assigned a continuous latent maturity variable
\begin{equation}
\label{eq:latent_maturity}
\ell_i = \tilde{\ell}_i + \epsilon_i,
\qquad
\tilde{\ell}_i \in \{1,2,3,4\},
\qquad
\epsilon_i \sim \mathcal{N}(0,\sigma^2)
\end{equation}
where $\tilde{\ell}_i$ denotes a discrete organizational governance maturity class and $\epsilon_i$ is a small stochastic perturbation. The base maturity level class $\tilde{\ell}_i$ is drawn from a categorical prior,
\begin{equation}
\label{eq:maturity_prior}
\mathbb{P}(\tilde{\ell}_i = k) = \pi_k,
\qquad
(\pi_1,\pi_2,\pi_3,\pi_4) = (0.40,\,0.30,\,0.20,\,0.10)
\end{equation}
reflecting the empirical scarcity of highly mature organizations. We use $\sigma = 0.25$ and clip perturbations to the valid maturity range, inducing correlation across practices while allowing local variability.

\paragraph{Modeling CSF practice difficulty.}
For each NIST CSF practice indexed by $p \in \{1,\dots,P\}$ (with $P = 42$), we define three ordered cut points $\mathbf{b}_p = (b_{p1}, b_{p2}, b_{p3})$ that encode the relative operational cost and complexity of achieving higher maturity tiers. We initialize all practices with baseline cut points $(1.5,\,2.5,\,3.5)$ and apply a practice-specific difficulty shift
\begin{equation}
\label{eq:difficulty_shift}
\Delta_p
=
0.3 \times
\left(
\frac{\text{Cost}_p + \text{Complexity}_p}{2}
- 2.5
\right)
\end{equation}
providing adjusted cut points $b_{pk} = b^{(0)}_k + \Delta_p$. Easier practices shift downward, while more demanding practices shift upward.

\paragraph{Sampling practice maturity tiers.}
Using the organization-level latent maturity $\ell_i$, the maturity tier for practice $p$ in organization $i$, denoted $t_{ip} \in \{1,2,3,4\}$, is sampled using an ordered logit model with logistic cumulative distribution function $\sigma(z) = 1/(1+e^{-z})$. We define the cumulative probabilities
\begin{equation}
\label{eq:ordered_logit_cdf}
F_{pk}(\ell_i)
=
\sigma\!\big(b_{pk} - \ell_i\big),
\qquad
k = 1,2,3
\end{equation}
from which the tier probabilities follow as
\begin{equation}
\label{eq:ordered_logit_category_probs}
\begin{aligned}
\pi_{p1}(\ell_i) &= F_{p1}(\ell_i), \\
\pi_{p2}(\ell_i) &= F_{p2}(\ell_i) - F_{p1}(\ell_i), \\
\pi_{p3}(\ell_i) &= F_{p3}(\ell_i) - F_{p2}(\ell_i), \\
\pi_{p4}(\ell_i) &= 1 - F_{p3}(\ell_i).
\end{aligned}
\end{equation}

We draw $t_{ip}$ by inverse CDF sampling. This formulation ensures that organizations with higher latent maturity are more likely to achieve higher tiers across all practices, preserving coherence while allowing variation in specific practices.

This construction yields three desirable properties: higher latent maturity increases the likelihood of higher tiers, the maturity prior biases populations toward lower tiers, and higher tiers remain rare for more complex practices.

\subsection{Synthetic Data Generation Procedure}

Algorithmically, given a table of CSF practice cut points and a selected set of practices:
\begin{enumerate}
  \item For each organization $i=1,\dots,N$, draw a base organizational maturity class 
  $\tilde{\ell}_i \sim \mathrm{Categorical}(\pi_1,\dots,\pi_4)$ and set
  $\ell_i = \tilde{\ell}_i + \epsilon_i$, where $\epsilon_i \sim \mathcal{N}(0,\sigma^2)$
  and $\ell_i$ is clipped to the valid maturity range.

  \item For each organization $i$ and practice $p$, compute the tier probabilities
  $(\pi_{p1}(\ell_i), \pi_{p2}(\ell_i), \pi_{p3}(\ell_i), \pi_{p4}(\ell_i))$.

  \item Sample the practice maturity tier $t_{ip} \in \{1,2,3,4\}$ by inverse CDF sampling.

  \item Emit a record
  $\{\texttt{org\_id}=i,\; p_1{:}t_{i1},\dots,p_P{:}t_{iP}\}$.
\end{enumerate}

The organizational maturity prior $(\pi_1,\dots,\pi_4)$ governs the relative frequency of maturity classes, while $\sigma$ controls heterogeneity within each class.

The resulting CSF practice maturity tiers are mapped to ATT\&CK mitigation maturity signals using the weighted scheme described in Section~\ref{subsec-csf-mapping}. We generate synthetic datasets of configurable size, where each record contains an \texttt{org\_id} and assigned maturity tiers for each practice.

\subsection{ATT\&CK Flow Corpus and Sequence Extraction}

We parse ATT\&CK Flow JSON bundles from the Center for Threat Informed Defense (CTID) corpus\footnote{\url{https://center-for-threat-informed-defense.github.io/attack-flow/}} and additional flows that we derived from open-source threat reports. Each attack flow encodes a documented cyber operation as a structured graph containing attack-technique nodes and relationship edges that represent causal links between steps in the operation. We convert each bundle into a directed multigraph, start from \texttt{start} edges, and follow \texttt{effect}, \texttt{asset}, and \texttt{object} links to enumerate root-to-leaf traversals. We drop any non-move leaves (e.g., \texttt{tool}, \texttt{attack-asset}). When a flow branches, each branch is treated as a distinct continuation, yielding multiple sequences from a single bundle.

Each attack action node is represented as a token consisting of tactic and technique:
\[
\texttt{token} \;=\; \texttt{TAxxxx}:\texttt{Tyyyy[.zzz]},
\]
e.g., \texttt{TA0001:T1190}. If a node includes a \texttt{certainty} field, we compute an unnormalized path score by multiplying certainty values along a depth-first traversal. For example, a path with certainties $0.9$, $0.8$, and $0.7$ yields an unnormalized score of $0.504$. This score reflects the relative confidence of the path within the flow but is not yet used directly for sampling.

To prevent highly branched flows from dominating the corpus, we normalize path scores \emph{within each bundle} so that the total weight across all extracted paths sums to one. After normalization, a bundle that fans out into four terminal paths assigns each path a sampling weight of $0.25$, regardless of its raw certainty product. These normalized weights are used when aggregating sequences for training to ensure that each documented operation contributes equally while preserving relative structure within the flow.

\subsection{Weighted VOMM Estimation}

We estimate weighted context counts:
\begin{equation}
\label{eq:weighted_context_count_app}
C(\mathbf{c}, a)
=
\sum_{n=1}^{N} w^{(n)}
\sum_{i=1}^{L_n}
\mathbb{I}[
(x_{i-k}, \dots, x_{i-1}) = \mathbf{c}
\wedge
x_i = a
]
\end{equation}

and apply smoothing with back-off:
\begin{equation}
\label{eq:vomm_backoff_app}
p(a \mid \mathbf{c})
=
\begin{cases}
\dfrac{C(\mathbf{c},a)+\alpha}{C(\mathbf{c})+\alpha |V|}, & C(\mathbf{c}) \ge m,\\
p(a \mid \text{backoff}(\mathbf{c})), & C(\mathbf{c}) < m,\\
|V|^{-1}, & \mathbf{c} = \varnothing.
\end{cases}
\end{equation}

\subsection{Example Transition}
Suppose the context $(\texttt{TA0001:T1190},\ \texttt{TA0006:T1136.001})$ is followed by:
\begin{itemize}
  \item \texttt{TA0003:T1059} with count $4$
  \item \texttt{TA0005:T1047} with count $1$
  \item all others $0$
\end{itemize}
With $\alpha = 1$ and $|V| = 200$, the smoothed probability for \texttt{T1059} is
\[
p(\texttt{T1059}\mid \mathbf{c}) = \frac{4+1}{(4+1)+200} = \frac{5}{205} \approx 0.024,
\]
while an unseen technique receives $1/205 \approx 0.0049$. This concentration encourages plausible next-step exploration without collapsing to a single deterministic continuation.

The VOMM encodes behavioral regularities (what attackers typically do next), and not success rates or exploit effectiveness. Its reliability therefore depends on coverage of the underlying flow corpus. Smoothing, back-off, and per-bundle normalization improve robustness under sparse and uneven reporting. Used as a prior alongside RL, it reduces unnecessary exploration and accelerates convergence while still allowing the learned policy to deviate when the environment reward indicates novel sequences.

\section{Base Budget Lookup Map}
\label{app:budget-map}

\begin{figure}[t]
  \centering
  \includegraphics[width=0.85\linewidth]{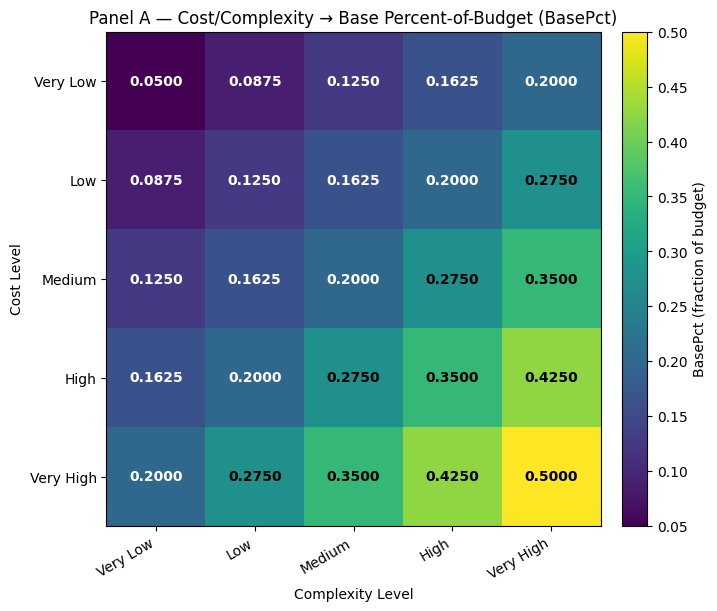}
  \caption{Base percent-of-budget lookup map ($\mathrm{PctCost}$) for ordinal cost and
  complexity ratings. Color intensity indicates the fraction of the episode budget
  consumed by a mitigation prior to maturity scaling.}
  \label{fig:basepct}
\end{figure}

Figure~\ref{fig:basepct} visualizes the base fraction of the episode budget associated with each combination of ordinal cost and complexity ratings. This lookup map is used by the $\mathrm{PctCost}(\cdot)$ function described in Section~\ref{subsec:reward} to translate qualitative mitigation characteristics into quantitative budget shares.

The map is monotone in both cost and complexity and has an upper bound of $0.5$, which ensures that even the most expensive mitigation consumes at most half of the episode budget prior to maturity scaling. This bound permits meaningful composition of mitigations within an episode while preserving clear tradeoffs for higher cost and complexity actions.

For a mitigation with cost level $c$ and complexity level $k$, the corresponding base
budget fraction $\mathrm{PctCost}(c,k)$ is read directly from the map. This value is
multiplied by the maturity scaling factor $\mu(m)$ to determine the final
episode cost of the mitigation. The scaling function is constructed so that higher
mitigation maturity reduces the budget less than lower mitigation maturity. Thus, differences in maturity drive the organizational capacity to execute mitigations.

\section{Adversary Resource Spread Model}
\label{app:resource-spread}
\begin{figure}[t]
  \centering
  \includegraphics[width=0.85\linewidth]{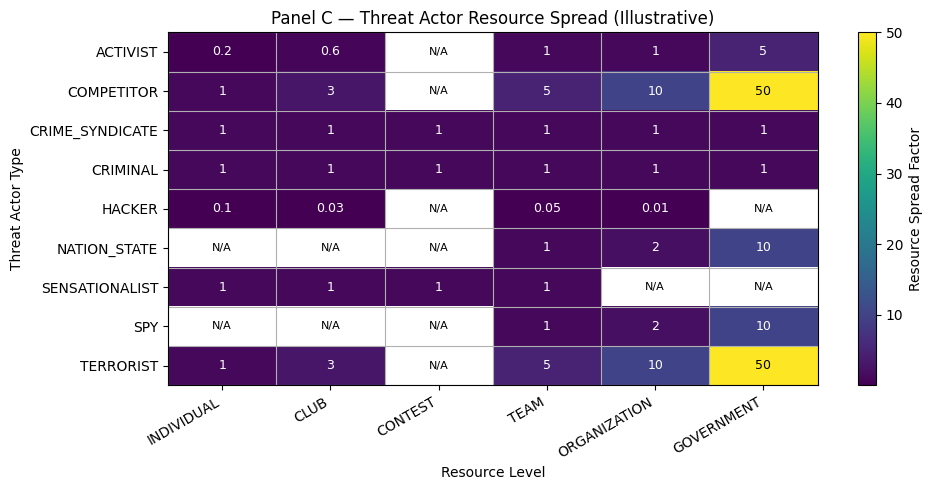}
  \caption{Resource-spread map $\mathrm{Spread}(\text{type},\text{resource})$ used to
  estimate per-target resource availability for adversaries. Values represent expected
  operator-equivalents allocated per target per planning period.}
  \label{fig:resource-spread}
\end{figure}

Figure~\ref{fig:resource-spread} summarizes the resource-spread model used to estimate how adversary resources are distributed across concurrent targets. The function $\mathrm{Spread}(\text{type},\text{resource})$ returns the expected per-target resource availability (in operator-equivalents per target per planning period) for adversary type and resource level.

Intuitively, adversaries with limited staffing and broad targeting must spread effort thinly across targets, yielding smaller per-target budgets, whereas adversaries with greater staffing and narrower targeting can allocate more effort per target. We compute $\mathrm{Spread}$ from analyst-provided ranges for (i) approximate adversary staffing and (ii) typical monthly targeting volume for that adversary type. The returned value is the ratio of these quantities, expressed per target:
\[
\mathrm{Spread} \;=\; \frac{\text{operators}}{\text{targets per period}}.
\]
For example, 10 operators targeting 1000 organizations per month yields
$\mathrm{Spread}=0.01$ operator-equivalents per target per month.

\section{Attack Path Reconstruction via Beam Search}
\label{app:beam_reconstruction}

\paragraph{Beam Search Algorithm.}
The reconstruction proceeds using a beam search of width $k$ over depth $d$. Let $\mathcal{B}_t$ denote the beam at depth $t$. The algorithm is defined as:

\begin{algorithm}[t]
\caption{Attack Path Reconstruction via Beam Search}
\label{alg:beam_reconstruction}
\begin{algorithmic}[1]
\Require Beam width $k$, maximum depth $d$, root observation $o_0$, initial state $s_0$
\Ensure Set of highest-scoring attack paths

\State $q_{\mathrm{root}} \gets Q(o_0,\cdot)$
\State $A_{\mathrm{root}} \gets \textproc{BuildRootPortfolio}(q_{\mathrm{root}}, \text{budget})$
\State $\mathcal{B}_0 \gets \{s_0\}$

\For{$t = 0$ to $d-1$}
    \State $\mathcal{C}_{t+1} \gets \emptyset$

    \ForAll{$s \in \mathcal{B}_t$}
        \If{\textproc{IsTerminal}($s$)}
            \State $\mathcal{C}_{t+1} \gets \mathcal{C}_{t+1} \cup \{s\}$
            \State \textbf{continue}
        \EndIf

        \State $\mathcal{T} \gets \textproc{VOMMCandidates}(s)$ \Comment{with observed-technique bias}

        \ForAll{$\tau \in \mathcal{T}$}
            \State $a \gets \textproc{SelectBestMitigation}(A_{\mathrm{root}}, \tau)$
            \State $(s', r, \tau) \gets \textproc{SimulateStep}(s, a, \tau)$
            \State $\mathcal{C}_{t+1} \gets \mathcal{C}_{t+1} \cup \{s'\}$
        \EndFor
    \EndFor

    \ForAll{$s' \in \mathcal{C}_{t+1}$}
        \State $\textproc{Score}(s') \gets R(s') + P(s') + U(s') + I(s') - \lambda D(s')$
    \EndFor

    \State $\mathcal{B}_{t+1} \gets$ Top-$k$ states in $\mathcal{C}_{t+1}$ by score
\EndFor

\State \Return highest-scoring paths across $\{\mathcal{B}_0, \dots, \mathcal{B}_d\}$
\end{algorithmic}
\end{algorithm}

A state $s$ is terminal if the adversary reaches an impact objective, exhausts its resource budget, or the maximum search depth is reached. Because each node contains a full environment state, this process captures both adversary dynamics and defender effects over time. 

Running this procedure independently for each prioritized adversary yields a set of $k$ high-scoring attack paths that are both (i) consistent with historical technique usage (via the VOMM) and (ii) aligned with a learned mitigation policy (via the DQN). Along these paths we observe which specific mitigations the defender repeatedly activates to disrupt that adversary’s progress. Aggregating across all adversaries produces a finite set of candidate mitigations, each annotated with (a) the adversaries and techniques it counters, (b) its estimated impact, and (c) its contribution to cumulative reward.